\documentclass[sigconf]{aamas} 

\usepackage{balance} 
\usepackage{array}
\usepackage{float}
\usepackage{mathtools}
\usepackage{multirow}
\usepackage{amsmath,amsfonts}
\usepackage{graphicx}
\usepackage{subfigure}
\usepackage[belowskip=-0.6pt,aboveskip=-0.4pt]{caption}
\newcommand\cparagraph[1]{\vspace{0.6mm}\noindent\textbf{#1.}}
\usepackage{tikz}
\def\checkmark{\tikz\fill[scale=0.4](0,.35) -- (.25,0) -- (1,.7) -- (.25,.15) -- cycle;}

\DeclareMathOperator{\EX}{\mathbb{E}}%

\setcopyright{ifaamas}
\acmConference[AAMAS '23]{Proc.\@ of the 22nd International Conference
on Autonomous Agents and Multiagent Systems (AAMAS 2023)}{May 29 -- June 2, 2023}
{London, United Kingdom}{A.~Ricci, W.~Yeoh, N.~Agmon, B.~An (eds.)}
\copyrightyear{2023}
\acmYear{2023}
\acmDOI{}
\acmPrice{}
\acmISBN{}

\acmSubmissionID{???}

\title[Hard Constraints in De-centralized Learning]{Discrete-choice Multi-agent Optimization: Decentralized Hard Constraint Satisfaction for Smart Cities}

\author{Srijoni Majumdar}
\affiliation{
\institution{University of Leeds, UK}
\city{}
\country{}}
\email{s.majumdar@leeds.ac.uk}

\author{Chuhao Qin}
\affiliation{
\institution{University of Leeds, UK}
\city{}
\country{}}
\email{sccq@leeds.ac.uk}

\author{Evangelos Pournaras}
\affiliation{
\institution{University of Leeds, UK}
\city{}
\country{}}
\email{e.pournaras@leeds.ac.uk}

\begin{abstract}
Making Smart Cities more sustainable, resilient and democratic is emerging as an endeavor of satisfying hard constraints, for instance meeting net-zero targets. Decentralized multi-agent methods for socio-technical optimization of large-scale complex infrastructures such as energy and transport networks are scalable and more privacy-preserving by design. However, they mainly focus on satisfying soft constraints to remain cost-effective. This paper introduces a new model for decentralized hard constraint satisfaction in discrete-choice combinatorial optimization problems. The model solves the cold start problem of partial information for coordination during initialization that can violate hard constraints. It also preserves a low-cost satisfaction of hard constraints in subsequent coordinated choices during which soft constraints optimization is performed. Strikingly, experimental results in real-world Smart City application scenarios demonstrate the required behavioral shift to preserve optimality when hard constraints are satisfied. These findings are significant for policymakers, system operators, designers and architects to create the missing social capital of running cities in more viable trajectories. 

\end{abstract}


\newcommand{\BibTeX}{\rm B\kern-.05em{\sc i\kern-.025em b}\kern-.08em\TeX}

\begin{document}


\pagestyle{fancy}
\fancyhead{}


\maketitle 


\section{Introduction}

\begin{figure}[!htb]
  \includegraphics[scale=0.2]{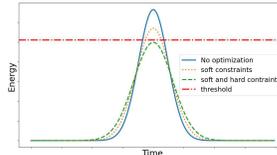}
   \caption{An illustrative optimization scenario. \emph{Baseline}: Without any optimized demand response, the power peak causes a blackout. \emph{Soft constraints}: Significant power peak reduction that does not though prevent the power blackout. \emph{Hard constraints}: Guarantee the reduction of the power peak below the blackout threshold.}
  \label{fig:intro1}
\end{figure}

Setting hard constraints in how we consume, produce, distribute and manage urban resources becomes paramount for the sustainability of our cities~\cite{Helbing2021}. Coordinated responses to climate change often aim to satisfy hard constraints for carbon footprint emissions and net-zero~\cite{Ramaswami2021}. Smart Grid technologies are still under development because of challenges to satisfy hard operational constraints that can cause catastrophic power blackouts~\cite{Pournaras2016} (see Figure~\ref{fig:intro1}). The satisfaction of hard constraints is also the safeguards for safety and the social capital for trust in establishing autonomous self-driving cards at scale~\cite{Du2021}. Currently, the complexity, scale and decentralization of socio-technical infrastructures in Smart Cities are a barrier for satisfying hard constraints by design. Instead, soft constraints prevail in the vast majority of optimization and learning approaches applied to the broader spectrum of Smart City applications~\cite{Khan2018,Niesse2016,pournaras2018decentralized,Pournaras2020}. This research gap is the focus and subject of this paper.

This paper studies the decentralized hard constraint satisfaction in discrete-choice multi-agent optimization, in particular distributed multi-objective combinatorial optimization problems. This is a large class of problems~\cite{Pournaras2020} in which agents autonomously determine a number of finite options to choose from (operational flexibility). The agents may have their own preferences over these alternatives, expressed with a \emph{discomfort cost} for each option. However, such choices often turn out to be inter-dependent to minimize system-wide \emph{inefficiency and unfairness costs} (non-linear cost functions) for which the agents may have (explicitly or implicitly) interest as well. These choices require coordination and computing the optimal combination of choices in an NP-hard problem~\cite{pournaras2018decentralized}. This is especially the case when agents come with different levels of \emph{selfish vs. altruistic} behavior, with which they prioritize the minimization of their individual discomfort cost over the collective inefficiency and unfairness cost. Smart Cities are full of emerging application scenarios that can be modelled as such optimization problems~\cite{pournaras2018decentralized,Pournaras2020,Qin2022}: power peak reduction to avoid blackouts, shift of power demand to consume more available renewable energy resources, coordinated vehicle routing to decrease travel times, traffic jams and air pollution, coordinated swarms of Unmanned Aerial Vehicles (UAVs) for distributed sensing, load-balancing of bike sharing stations, and other. 

So far, heuristics for solving these distributed multi-agent optimization problems mainly address soft constraints, which is the best effort to minimize all involved costs. This is because in the absence of complete information in the agents, it is simpler to design algorithms that search efficiently for good solutions even if these are not the optimal ones. Instead, satisfying hard constraints opens up a Pandora box: without full information, any autonomous agent choice can violate the hard constraints that can only be satisfied with certainty at an aggregate level. Letting agents make conservative choices to avoid violating the hard constraints may significantly downgrade performance and optimality, while any rollback of choices violating these constraints is complex and costly.

To address this timely problem with impact on Smart Cities, a new decentralized hard constraint satisfaction model is introduced. The model constructs ranges of upper and lower bounds within which the aggregate choices and costs must remain, while optimizing for soft constraints. To solve the cold start problem in the initialization phase during which agents choices and costs are undergoing aggregation, a heuristic is introduced for the agents to make choices with the highest average likelihood to satisfy all hard constraints. As this heuristic is sensitive to the agents' self-organization (order) in decision-making and aggregation, agents keep reorganizing themselves after violations of hard constraints as long as a stopping criterion is not reached. After the cold start phase and after agents successfully satisfy the hard constraints, they can shift entirely to the optimization of the soft constraints while preserving the satisfaction of the hard constraints locally  with a low cost.

The proposed model is integrated into a collective learning algorithm, the \emph{Iterative Economic Planning and Optimized Selections} (I-EPOS)~\cite{pournaras2018decentralized}. This allows a comprehensive assessment of the decentralized hard constraint satisfaction model and its impact on the optimality of the soft constraints. For the first time, experimental evaluation with real-world data from Smart City scenarios disentangle the performance sacrifice as a result of satisfying hard constraints and the additional agents' altruism level required to mitigate such sacrifice. These findings are highly revealing for system operators, policymakers, system designers and architects. They can inform them about the additional social capital (incentives/rewards) that they require to build (and pay) to preserve the cost-effectiveness of socio-technical infrastructures operating with hard constraints.

The contributions of this paper are summarized as follows: (i) A model of decentralized hard constraint satisfaction on optimizing aggregate agents' choices and their aggregate costs. (ii) The instantiation of this model on a decentralized multi-objective combinatorial optimization algorithm of collective learning for multi-agent systems. (iii) The applicability of decentralized hard constraint satisfaction on three Smart Cities scenarios using real-world data: energy, bike sharing and UAVs swarm sensing. (iv) Insights about the optimality sacrifice as moving from soft to hard constraints and how this optimality loss is measured in terms of the required behavioral shift to preserve performance, i.e. restoring altruism deficit. (v) An open-source software artifact implementation of the model for the I-EPOS collective learning algorithm.

This paper is summarized as follows: Section~\ref{sec:related-work} reviews related methods. Section~\ref{sec:Methodology} introduces the decentralized hard constraint satisfaction model. Section~\ref{sec:imple} illustrates the applicability of this model to the collective learning algorithm of I-EPOS and its implementation. Section~\ref{sec:Results} illustrates the experimental methodology and the evaluation.  Section~\ref{sec:Conclusion} concludes this paper and outlines future work.

\begin{table*}[!htb]
\centering
\begin{small}
\caption{Comparison of self adaptive decentralized approaches}
\label{tab:compare1}
\begin{tabular}{p{3.6cm}p{3.1cm}p{3.1cm}p{3.1cm}p{3.1cm}}
\toprule
Attributes	&	I-EPOS~\cite{pournaras2018decentralized}	&	EPOS~\cite{Pournaras2013} & COHDA~\cite{hinrichs2014cohda,niesse2016local,niesse2014conjoint}  & H-DPOP~\cite{chen2018bd,kumar2008h}	\\ \hline
Plan Selection -
Autonomy
& Locally & Parent-Level & Locally & Parent-Level\\ \hline
Computational Cost &
agent: $O(pt)$ ; system: $O(pt\ log\ a)$
&
agent: $O(p^c)$ ; system: $O(p^c\ log\ a)$
&
agent: $O(pt)$ ; system: $O(pt)$  &  agent: $O(p^c)$ ; system: $O(p^c)$ \\ \hline

 Communication Cost &
agent: $O(t)$ ; system: $O(t\ log\ a)$
&
agent: $O(p)$ ; system: $O(p\ log\ a)$
&
agent: $O(at)$ ; system: $O(at)$  &  agent: $O(p^c)$ ; system: $O(p^c)$ \\ \hline
Information Exchange & tree overlay; aggregate information;  & tree overlay; aggregate messages;  & k-connected graph; full information;  
 & tree overlay; full information;
\\ \hline
Soft Constraints & local (initialization), aggregated choices in global plan  & local (initialization), aggregated choices in global plan  & global  &  no soft constraints\\ \hline

Hard Constraints & local (initialization) & local (initialization) & local (initialization), global  & local (initialization), global \\ \hline
\multicolumn{5}{p{15cm}}{ $p$: number of plans (options), $t$: number of iterations, $c$: number of children, $a$: number of agents}

\end{tabular}
\end{small}
\end{table*}

\section{Comparison to Related Work}\label{sec:related-work}



We study a discrete-choice distributed multi-objective optimization problem for multi-agent systems with both soft and hard constraints. In such systems, most related optimization approaches~\cite{kaddoum2011optimization} operate as partially observable systems with agents communicating with their neighbors.

 There are approaches that rely on an asynchronous hierarchical process using depth-first search to order agents that communicate with their parents to make choices and optimize objective functions~\cite{billiau2012sbdo}. Mailler et al.~\cite{mailler2004solving} cluster agents based on the constraints they attempt to satisfy, with a central controller that uses a branch and bound paradigm for searching solutions. These hierarchical approaches suffer from failure risks, performance bottlenecks, and potential privacy breaches in application scenarios involving sensitive personal data, e.g. location and health data.

Multi-agent reinforcement learning approaches with constraints on agent choices are earlier studied. For instance, Curran et al.~\cite{curran2013addressing} generate rewards for agents to optimize delay intervals that prevent air traffic congestion with greedy scheduling to implement {\em hard stop} (constraints) for agents when the delay surpasses a limit.  Rollback to previous states ({\em warm restarts}) are earlier studied upon violation of global hard constraints in~\cite{parnika2021attention}. Simao et al.~\cite{simao2021alwayssafe} learn non-violated execution by training using datasets containing constrained actions of the agents and corresponding global states of a centrally controlled environment. Even though the execution is decentralized, the learned model used to provide recommendations to the agents is an outcome of a centralized computation.

Decentralized and asynchronous versions of population search-based optimization methods, such as particle swarm optimization~\cite{akat2008decentralized} or ant colony~\cite{gupta2020comparative} algorithms show a slow convergence with high communication cost to rollback after violations of hard constraints, while improving global fitness and local search. This may slow down online real-time adaptations. 
Violation of hard constraints is prevented via message broadcasting that rolls back all choices made after a violation~\cite{billiau2012sbdo}.

Other earlier approaches use tree overlay network structures for aggregation to aggregate messages  from child nodes in form of hypercubes to reduce the frequency of message exchanges~\cite{deng2019pt,chen2018bd,kumar2008h}. 
These methods use dynamic programming approach and thus storing all solutions increases the size of messages  exponentially.


Other highly efficient discrete-choice multi-agent optimization methods, such as COHDA~\cite{hinrichs2014cohda} and EPOS~\cite{pournaras2018decentralized}, address a large spectrum of NP-hard combinatorial problems in the domains of Smart Grids and Smart Cities~\cite{Niesse2016,hinrichs2017distributed,Pournaras2020,pournaras2017self}. COHDA generalizes well in different communication structures among the agents that have full view of the systems, while EPOS focuses on hierarchical acyclic graphs such as trees to perform a cost-effective decision-making and aggregation of choices. Table~\ref{tab:compare1} compares the design and efficiency of multi-agent optimization approaches, as well as how they address soft and hard constraints.

As COHDA shares full information between agents, it has higher communication cost. The computational cost is lower at global level for COHDA compared to EPOS because of the agents' brute force search to aggregate choices. 
Both COHDA and EPOS focus on satisfying soft constraints, like minimizing cost functions that satisfy balancing (minimum variance and standard deviation~\cite{pournaras2018decentralized}) or matching  (minimum root mean square error, residual sum of squares~\cite{pournaras2017self}) objectives. However, satisfying global hard constraints (Table~\ref{tab:compare1}) is challenging as agents need to additionally coordinate for choices, whose potential violations are only confirmed at an aggregate level, which makes any rollback of choices to avoid violations particularly complex.  An expensive rollback procedure earlier introduced in COHDA~\cite{niesse2014conjoint} performs complete rescheduling using a 0-1 multiple-choice combinatorial to find another solution that satisfy the constraints.

Summarizing, satisfaction of  hard constraints during initialization phase, when agents accumulate information about other agents' choices (cold start problem) remains an open challenge. It is also unclear how the satisfaction of hard constraints degrades the performance of these efficient algorithms based on their soft constraints. Addressing these open questions is the focus  of this paper.

\section{Hard Constraint Satisfaction Model}\label{sec:Methodology}
This section introduces the general optimization problem and the decentralized hard satisfaction model.

\subsection{Optimization problem}\label{subsec:problem}

Table~\ref{tab:notation} summarizes the mathematical symbols of this article. Assume a socio-technical systems of $n$ users, each assisted by a software agent, i.e. a personal digital assistant. Each agent $i$ has a number of $k$ options to choose from. Each option $j$ is referred to as a \emph{possible plan}, which is a sequence of real values $p_{i,j}=(p_{i,j,u})_{u=1}^{m} \in P_{i}=(p_{i,j})_{j=1}^{k}, \forall i \in \{1,...,n\}$. Each agent selects one and only one plan $p_{i,s}$, which is referred to as the selected plan (i.e. the agent's choice). All agents' choices aggregate element-wise to the collective choice, the \emph{global plan} $g=(g_{u})_{u=1}^{m}=\sum_{i=1}^{n}p_{i,s}$ of the multi-agent system. A possible plan of an agent may represent a resource schedule or allocation, e.g. the energy consumed over time or the energy consumed from a certain supplier. Multiple possible plans for each agent represent alternatives and its operational flexibility. In the example of energy, the global plan represents the total energy consumption in the system over time or suppliers (see Figure~\ref{fig:intro1}).
\begin{table}[!htb]
\centering
\begin{footnotesize}
\caption{Mathematical notations used in this paper.}
\label{tab:notation}
\begin{tabular}{p{2cm}p{5.8cm}}
\toprule
Notation   &    Meaning   \\ \hline

$n$ & Number of agents\\ \hline

$P_i$ & Set of possible plans for agent $i$ \\  \hline

$m$ & Plan size \\ \hline

$k$ & Number of plans \\ \hline

$p_{i,j} \subset \mathbb{R}^m$ & The $j^{th}$ plan as sequence of $m$ elements of agent $i$\\  \hline

$p_{i,s}$ & Selected plan of agent $i$ \\  \hline

$g=\sum_{i=1}^{n}p_{i,s}$ & Global plan from selected plans of all $n$ agents\\  \hline

$\beta_i$ & Discomfort factor for  agent $i$\\  \hline

$\alpha_i$ & Unfairness factor for agent $i$\\  \hline
$r$ & Constraints satisfaction rate \\ \hline

$I$ & Inefficiency cost \\ \hline
$D$ & Discomfort cost \\ \hline
$U$ & Unfairness cost \\ \hline
$f_{\mathsf{D}}:\mathbb{R}^m \rightarrow \mathbb{R}$ & Discomfort cost function \\  \hline

$f_{\mathsf{I}}:\mathbb{R}^m \rightarrow \mathbb{R}$ & Inefficiency cost function \\  \hline

$f_{\mathsf{U}}:\mathbb{R}^m \rightarrow \mathbb{R}$ & Unfairness cost function\\  \hline

$\mathcal{U} \subset \mathbb{R}^m$ & Sequence of upper bound hard constraints \\  \hline

$\mathcal{L} \subset \mathbb{R}^m$ & Sequence of lower bound hard constraints \\  \hline

$\EX(p_{i,j},\mathcal{U})$ & Expected satisfaction for upper bound constraints \\  \hline

$\EX(p_{i,j},\mathcal{L})$ & Expected satisfaction for lower bound constraints \\  \hline

\end{tabular}
\end{footnotesize}
\end{table}

Agents' choices are made based on different, often opposing criteria. Each agent has its individual preferences over the possible plans, measured by the \emph{discomfort cost} $f_{\mathsf{D}}(p_{i,j})=D_{i,j}$ of each plan $j$, which also makes the mean discomfort cost in the system $f_{\mathsf{D}}(p_{1,s},...,p_{n,s})=\frac{1}{n}\cdot\sum_{i=1}^{n}f_{\mathsf{D}}(p_{i,s})$. Each agent can make independent choices to minimize their own discomfort cost. However, agents may also have interest to satisfy the following two general-purpose collective criteria: inefficiency cost $f_{\mathsf{I}}(\sum_{i=1}^{n}p_{i,s})=I_{i}$ and unfairness cost $f_{\mathsf{U}}(D_{1,s},...,D_{n,s})=U_{i}$. If these cost functions are non-linear, meaning the choices of the agents depend on each other, the satisfaction of soft constraints, i.e. minimizing the inefficiency and unfairness cost, is a combinatorial NP-hard optimization problem~\cite{pournaras2018decentralized}. Balancing (e.g. min variance) and matching objectives (e.g. min residual of sum squares) are examples for measuring inefficiency cost with a broad applicability in load-balancing application scenarios of Smart Cities: minimizing power peaks, shifting demand to times with high availability of renewable energy resources, rerouting vehicles to avoid traffic jams, etc.  Table~\ref{tab:ex} show such a case, in which three agents have two options (plans). These plans may represent energy consumption choices while forming an optimal global plans that meets the available energy supply. The elements may signify the power consumption for the day and night. The agents choose plans with minimum dispersion between elements (soft constraints), but that leads to a suboptimal global plan of [7,13]. The global plan should also come with lower dispersion. The variance of the discomfort costs over the population of agents can measure the unfairness cost. Satisfying all of these (opposing) objectives depends on the selfish vs. altruistic behavior of the agents, e.g. whether they accept a plan with a bit higher discomfort cost to decrease inefficiency or unfairness cost. We can observe this in Table~\ref{tab:ex}. If Agent C selects a plan ([6,2]) with higher energy requirement during the day, it achieves to minimize the inefficiency cost and an optimum global plan of [10,10] is achieved. Such multi-objective trade-offs are modelled with the parameters $\alpha_{i}$ and $\beta_{i}$ such that:
\setlength{\belowdisplayskip}{0pt} \setlength{\belowdisplayshortskip}{0pt}
\setlength{\abovedisplayskip}{0pt} \setlength{\abovedisplayshortskip}{0pt}
\begin{equation}
    \begin{split}
        p_{i,s}=\overset{k}{\underset{j=1}{argmin}}
        & (1-\alpha-\beta) \cdot f_{\mathsf{I}}(p_{1,s}+...+p_{i,j}+...+p_{n,s})\\
        & + \alpha \cdot f_{\mathsf{U}}(D_{1,s},...,D_{i,j},...,D_{n,s})\\
        & + \beta \cdot f_{\mathsf{D}}(D_{1,s},...,D_{i,j},...,D_{n,s}).
    \end{split}
    \label{eq:pqt}
\end{equation}

From the above equation, it is apparent that the choice of a plan cannot be easily optimized without (i) information of the other agents' choices and (ii) coordination of the agents' choices for non-linear cost functions that depend on each other. The  optimization heuristics (Section~\ref{sec:related-work}) address the satisfaction of such soft constraints via various sequential information exchange, information aggregation and communication schemes that coordinate agents' choices. See the baseline scenario of soft constraints in Table~\ref{tab:ex}.

\begin{table*}[!htb]
\caption{An example of a discrete-choice combinatorial optimization problem with three agents ($n=3$), each with two plans ($k=2, m=2$). All combinations of possible plan selections make $2^{3}=8$ possible global plans. Hard constraints with an upper bound on the aggregate choices (global plan $g$) are introduced with an expected satisfaction of $\sum_{u = 1}^{m} (\mathcal{U}_u - p_{i,j,u})$. (1) The baseline scenario is the soft constraints that minimize the inefficiency cost $f_{\mathsf{I}}(g) \approx |p_{i,j,1} - p_{i,j,2}|$. The global plan $[10,10]$ is the one with the inefficiency cost. (2) Agents choose plans that maximize the expected satisfaction. This results in the global plan of $[6,15]$ that satisfies the hard constraint $\mathcal{U}=[9,]$. (3) Similarly, the hard constraint $\mathcal{U}=[9,]$ is satisfied with the global plan of $[14,9]$. (4) Both new hard constraints $\mathcal{U}=[10,13]$ are satisfied with the global plan $[7,13]$. (5) The second hard constraint $\mathcal{U}=[9,9]$ is violated by the selected global plan $[7,13]$.  {\large $\surd$}: constraint satisfaction;  {\large $\times$}:  constraint violation in $p_{i,j}$ and $g$} 
\centering
\begin{tiny}
\renewcommand\arraystretch{1.4}
\begin{tabular}{p{2.9cm}|c|c|c|c|c|c|p{0.57cm}|p{0.57cm}|p{0.57cm}|p{0.57cm}|p{0.57cm}|p{0.57cm}|p{0.57cm}|p{0.57cm}}
\hline


\multicolumn{1}{p{1.17cm}|}{\bf Constraints} & \multicolumn{2}{c|}{\bf Agent A } & \multicolumn{2}{c|}{\bf Agent B} & \multicolumn{2}{c|}{\bf  Agent C}  & \multicolumn{8}{c}{\bf All Possible Global Responses} \\ \hline

 Agents' Plans ($p$) & {\bf [3, 5]} & {\bf [2, 7]} & {\bf [1, 3]}  & {\bf [5, 2]}  & {\bf [6, 2]}  & {\bf [3,5]}  & {\bf [10,10]} & {\bf  [14,9]} & {\bf [7,13]} & {\bf [11,12]} & {\bf [9,12]} & {\bf [13,11]} & {\bf [6,15]} & {\bf [10,14]}  \\ \hline

\multicolumn{15}{c} {\bf \footnotesize 1. Soft Constraints (Baseline) } \\ \hline

{\bf  Minimize Inefficiency Cost}  & |3-5| = 2 & |2-7| = 5 & |1-3| = 2 & |5-2| = 3 & |6-2| = 4 & |3-5| = 2 & & & & & & & & \\
{\bf Selected Plans} & {\checkmark} &  & {\checkmark} & & {\checkmark} &  & & & & & & & & \\
{\bf Selected Global Plan} & &  &  & & &  & {\checkmark} & &  & & & & & \\ \hline 

\multicolumn{15}{c} {\bf \footnotesize 2. Hard Constraints - $\mathcal{U}$ = [9, ] } \\ \hline
{\bf Maximize Expected satisfaction}  & (9-3)+0=6  & (9-2)+0=7 & (9-2)+0=7 & (9-5)+0=4 & (9-6)+0=3 & (9-3)+0=6 & & & & & & & & \\
{\bf Selected Plans} &  & {\checkmark} & {\checkmark} &  &  & {\checkmark} & &  & & & & & & \\ 
{\bf Selected Global Plan } &  &  &  &  &  & & &  & & & & & {\checkmark} & \\ \hline 
\multicolumn{15}{c} {\bf \footnotesize 3. Hard Constraints - $\mathcal{U}$ = [, 9] } \\ \hline
{\bf Maximize expected satisfaction } & 0+(9-5)=4  & 0+(9-7)=2 & 0+(9-3)=6 & 0+(9-2)=7 & 0+(9-2)=7 & 0+(9-5)=4 & & & & & & & & \\
{\bf Selected Plans} & {\checkmark} &  &  & {\checkmark} & {\checkmark}  & & &  & & & & & & \\ 

{\bf Selected Global Plan } &  &  &  & &  &  & & {\checkmark} & & & & & & \\ \hline 

\multicolumn{15}{c} {\bf \footnotesize 4. Hard Constraints - $\mathcal{U}$ = [10,13] } \\ \hline

{\bf  Maximize expected satisfaction }  & (10 - 3)+  & (10 - 2)+ & (10 - 1)+ & (10 - 5)+ & (10 - 6)+ & (10 - 3)+ & & & & & & & & \\

 & (13 - 5)=15   & (13 - 7)=14 & (13 - 3)=19  & (13 - 2)=17 & (13 -2)=15 & (13 - 5)=15 & & & & & & & & \\
{\bf Selected Plans} & {\checkmark} &  & {\checkmark} &  &  & {\checkmark} & &  & & & & & & \\ 

{\bf Selected Global Plan 
}  &  &  &  & &  & & &  & {\checkmark} & & & & & \\ \hline 

\multicolumn{15}{c} {\bf \footnotesize 5. Hard Constraints - $\mathcal{U}$ = [9,9]} \\ \hline
{\bf  Maximize expected satisfaction } & (9 - 3)+  & (9 - 2)+ & (9 - 1)+ & (9 - 5)+ & (9 - 6)+ & (9 - 3)+ & & & & & & & & \\

 & (9 - 5)=10  & (9 - 7)=9 & (9 - 3)=14  & (9 - 2)=11 & (9 - 2)=10 & (9 - 5)=10 & & & & & & & & \\
{\bf Selected Plans} & {\checkmark} &  & {\checkmark} &  &  & {\checkmark} & &  & & & & & & \\ 
{\bf Selected Global Plan} &  &  &  & &  & & &  & {\Large$\times$} & & & & & \\ \hline

\end{tabular}
\label{tab:ex}
\end{tiny}
\end{table*}

However, introducing hard constraints on the aggregated choices $g$ and their costs $D_{i,j}$, $I_{i}$, $U_{i}$ is challenging. This is because there is no guarantee to satisfy the hard constraints in the absence of full information, which is usually the case for decentralized heuristics that require time to converge to full information. This is a particular cold start problem of initialization/exploration, during which the first choices are made under high uncertainty. As choices with high likelihood of violating hard constraints add up incrementally, it becomes increasingly hard to discover choices that will prevent such violations. Hence, agents need different and more conservative selection criteria that prioritize hard over soft constraints. Designing and evaluating these criteria is a contribution of this paper.


\setlength{\belowdisplayskip}{-0.2pt} \setlength{\belowdisplayshortskip}{-0.2pt}
\setlength{\abovedisplayskip}{0pt} \setlength{\abovedisplayshortskip}{0pt}

\subsection{A heuristic for satisfying hard constraints}

The heuristic for satisfying the hard constraints on aggregate choices and their costs is illustrated in this section. Table~\ref{tab:ex} also illustrates an example of applying the heuristic in practice.
\vspace{-0.2cm}
\subsubsection{Constraints on aggregate choices}

For each element $g_{u}$ of a global plan $g$, a hard constraint is defined by a range (envelope) of an upper $\mathcal{U}=(\mathcal{U}_{u})_{u=1}^{m}$ and lower $\mathcal{L}=(\mathcal{L}_{u})_{u=1}^{m}$ bound, where $\mathcal{U}$, $\mathcal{L}$ are also sequences of real values of equal size $|\mathcal{U}|=|\mathcal{L}|=|g|=m$. Each value $u$ of the upper bound denotes that $\mathcal{U}_{u} \geq g_{u}$, whereas for the lower bound it holds that $\mathcal{L}_{u} \leq g_{u}$. 

The selected plan expected to satisfy all hard constraints at the initialization phase, during which the aggregate choices (global plan $g$) are not known, is estimated as follows:
\setlength{\belowdisplayskip}{0pt} \setlength{\belowdisplayshortskip}{0pt}
\setlength{\abovedisplayskip}{0pt} \setlength{\abovedisplayshortskip}{0pt}
\begin{equation}
p_{i,s} = \underset{p_{i,j} \in P_i}{argmax} \EX(p_{i,j},\mathcal{U},\mathcal{L}),
\end{equation}

\noindent where the expectation of satisfaction is given by:
\setlength{\belowdisplayskip}{0pt} \setlength{\belowdisplayshortskip}{0pt}
\setlength{\abovedisplayskip}{0pt} \setlength{\abovedisplayshortskip}{0pt}
\begin{equation}
\EX(p_{i,j},\mathcal{U},\mathcal{L}) = \sum_{u = 1}^{m} (\mathcal{U}_u - p_{i,j,u}) + \sum_{u = 1}^{m} (p_{i,j,u}- \mathcal{L}_{u}).
\end{equation}
\subsubsection{Constraints on aggregate costs}
The modeling for the hard constraints on the aggregate costs is exactly the same as the one of the aggregate choices, where the expected satisfaction for each of the costs of $f_{\mathsf{D}}(p_{1,s},...,p_{n,s})$, $f_{\mathsf{I}}(\sum_{i=1}^{n}p_{i,s})$ and $f_{\mathsf{U}}(D_{1,s},...,D_{n,s})$ is calculated for upper and lower bounds with $|\mathcal{U}|=|\mathcal{L}|=1$.



 

\subsubsection{Constraint satisfaction rate}

The effectiveness of the hard constraint satisfaction heuristic is measured by the satisfaction rate ($r$). This is the total number of satisfactions achieved out of a total number of trials made. These trials are often existing parameters of the optimization algorithms, for instance, random repetitions, or the order with which agents aggregate choices made to coordinate and optimize their own choices.



\section{Hard Constraints Implementation}\label{sec:imple}
The model of decentralized hard constraint satisfaction is implemented and integrated into the I-EPOS collective learning algorithm\footnote{Available at: https://github.com/epournaras/EPOS/tree/hard\_constraints.}~\cite{pournaras2018decentralized}.  I-EPOS solves a large class of optimization problems, as formalized in Section~\ref{subsec:problem}. It is chosen due to its large spectrum of applicability in Smart City scenarios~\cite{Pournaras2020} as well as its superior performance in satisfying soft constraints~\cite{pournaras2018decentralized}. Efficient coordinated choices are made using a self-organized~\cite{Pournaras2013} tree topology within which agents organize their interactions, information exchange and decision-making. I-EPOS benefits from the fact that trees are acyclic graphs: communication cost is very low and all exchanges in I-EPOS are at an aggregate level without double-counting. The coordination is a result of a more informed decision-making: each agent makes a choice taking into account the choices of a group of other agents (the tree branch underneath during initialization) or the choices of all agents at a previous time point (after initialization). Coordination evolves in multiple learning iterations, each consisting of a \emph{bottom-up} and \emph{top-down} phase. During the bottom-up phase, each agent chooses based on the new choices of the agents below and the choices of all agents in the previous iteration. However, each agent has an information gap: It has no information about the subsequent choices of the agents above in the tree. This problem is solved during the top-down phase, in which agents roll back (back propagation) to choices of the previous iteration as long as no costs reduction is achieved. Further information about the design of the I-EPOS collective learning algorithm is out of the scope of this paper and can be found in earlier work~\cite{pournaras2018decentralized}.

The decentralized hard constraint satisfaction model is implemented by filtering out the possible plans in Equation~1 that violate the given upper and lower bounds. However, in the first learning iteration, it is not possible determine these plans that violate the hard constraints with certainty because each agent only knows about the aggregate choices of the agents underneath (and not the ones above). As a result, the root agent may end up having no possible plan that does not violate the hard constraints. To prevent the likelihood of these violations, the agents make more conservative choices according to Equation~2 during the first iteration, aiming at maximizing the expected satisfaction of the hard constraints. Once the hard satisfactions are satisfied, the agents switch back to plan selection according to Equation~1, while keep filtering plans that violate the hard constraints. The agents cannot violate the hard constraints in these subsequent learning iterations because they always have the option to roll back to the choices made at the end of the first learning iteration during which hard constraints are satisfied (via the top-down phase).

\begin{figure}[!htb]
 \includegraphics[scale=0.3]{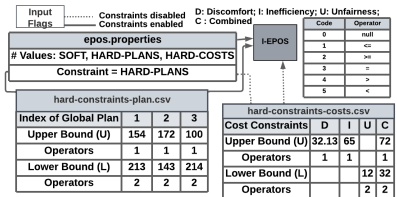}
  \caption{Implementation of decentralized hard constraints satisfaction in I-EPOS~\cite{majumdar_srijoni_2023_7791326}.}
 \label{fig:archi}
\end{figure}

Figure~\ref{fig:archi} illustrates the implementation of the hard constraints model on the open-source I-EPOS software artifact~\cite{majumdar_srijoni_2023_7791326}. The implementation of the cost function interfaces is extended to filter out plans that violate the hard constraints, as well as the selection based on the expected satisfaction principle of Equation~2. The hard constraints are controlled via the main input parameter file of I-EPOS (Java Properties). Constraints on aggregate choices and costs can be activated and deactivated. Two input .csv files are introduced, one for each type of hard constraints. Both contain the upper/lower bounds and the coding of the operators. 

\section{Experimental Evaluation}\label{sec:Results}

Table~\ref{tab:dataset} illustrates the application scenarios and settings of the experimental evaluation. A number of 1000 agents run the I-EPOS collective learning algorithm~\cite{pournaras2018decentralized}. They are self-organized~\cite{Pournaras2013} in a height-balanced
binary tree. The algorithm repeats 200 times, each time with a different random positioning of the agents in the tree. This introduces different decision-making order with which agents coordinate their optimized choices. At each repetition, the algorithm runs for 40 iterations, which is usually sufficient for convergence~\cite{pournaras2018decentralized}.  Evaluation is performed in three Smart City application scenarios: (i) energy demand-response, (ii) bike sharing and (iii) UAV swarm sensing. The optimized inefficiency cost function and the generation of plans are also outlined in Table~\ref{tab:dataset}.

\begin{table}[!htb]
\centering
\begin{footnotesize}

\caption{Parameters of the EPOS algorithm~\cite{pournaras2018decentralized} for datasets}
\label{tab:dataset}
\begin{tabular}{p{2.5cm}p{1.6cm}p{1.6cm}p{1.6cm}}
\toprule
Parameter	&	Energy 	&	Bike Sharing  & UAV Swarm\\ \hline
Num. of agents ($n$)
& 1000 & 1000 & 1000 \\ 

Num. of plans ($k$)
& 10 & 1 to 24 & 64\\

Plans size ($m$)
& 144 & 98 & 64 \\ 

Num. of repetitions
& 200 & 200 & 200 \\

Num. of iterations
& 40 & 40 & 40 \\ 

Inefficiency cost  $f_{\mathsf{I}}$
& Min VAR & Min VAR & Min RMSE\\
\bottomrule
\end{tabular}
\end{footnotesize}
\end{table}


\subsection{Smart City application scenarios}

\begin{figure*}[!htb]
    \centering
    \subfigure[$\mathcal{U}_{u}=1388$, $\mathcal{L}_{u}=1386$, $r=0.655$, $I=0.1006$.]{
                \includegraphics[height=2.8cm,width=4cm]{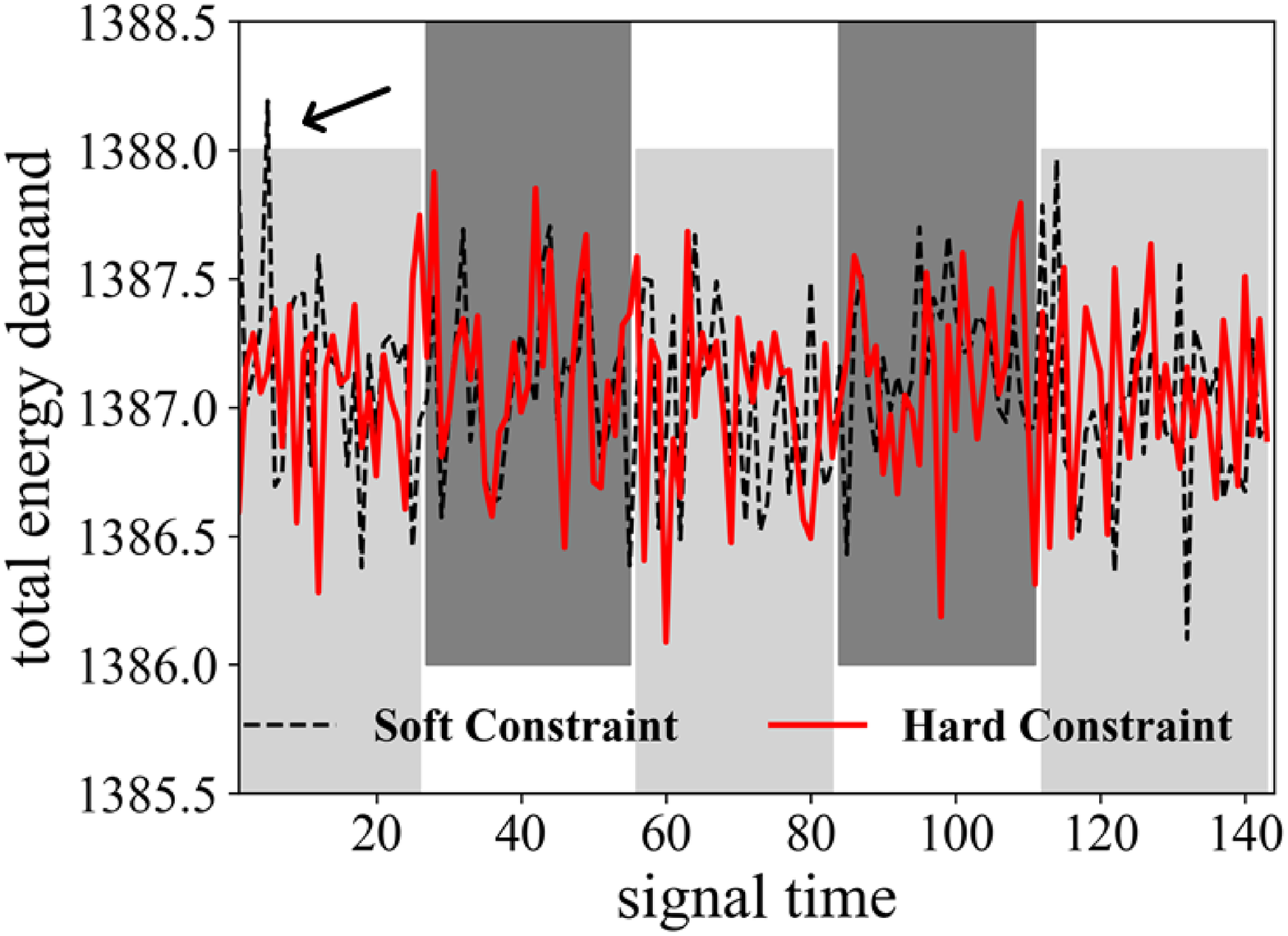}
                \label{fig:energy_level1}
        }
    \hspace{0.2cm}
    \subfigure[$\mathcal{U}_{u}=1387.7$, $\mathcal{L}_{u}=1386.2$, $r=0.335$, $I=0.1032$.]{
                \includegraphics[height=2.8cm,width=4cm]{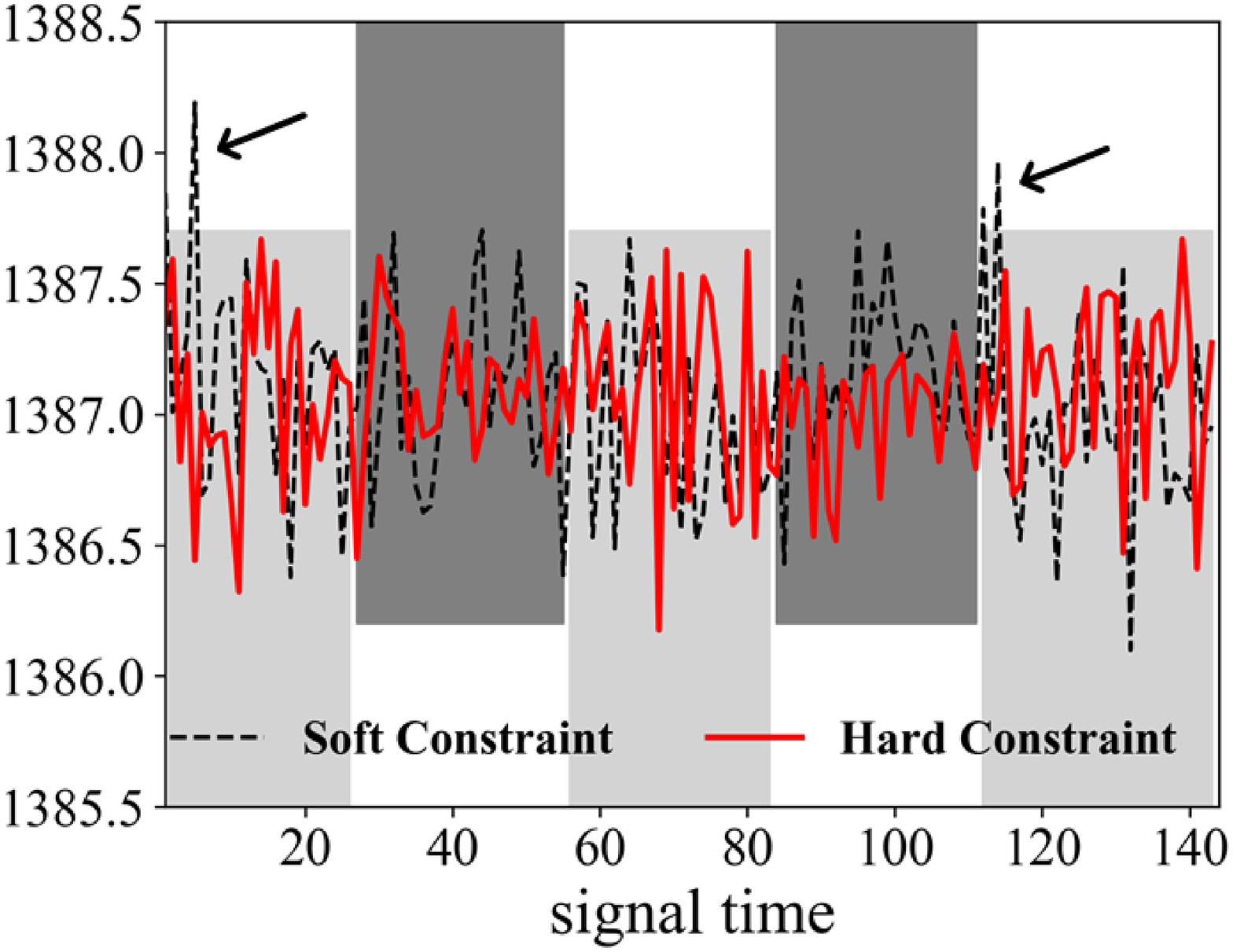}
                \label{fig:energy_level2}
        }
    \hspace{0.2cm}
    \subfigure[$\mathcal{U}_{u}=1387.6$, $\mathcal{L}_{u}=1386.4$, $r=0.07$, $I=0.1051$.]{
                \includegraphics[height=2.8cm,width=4cm]{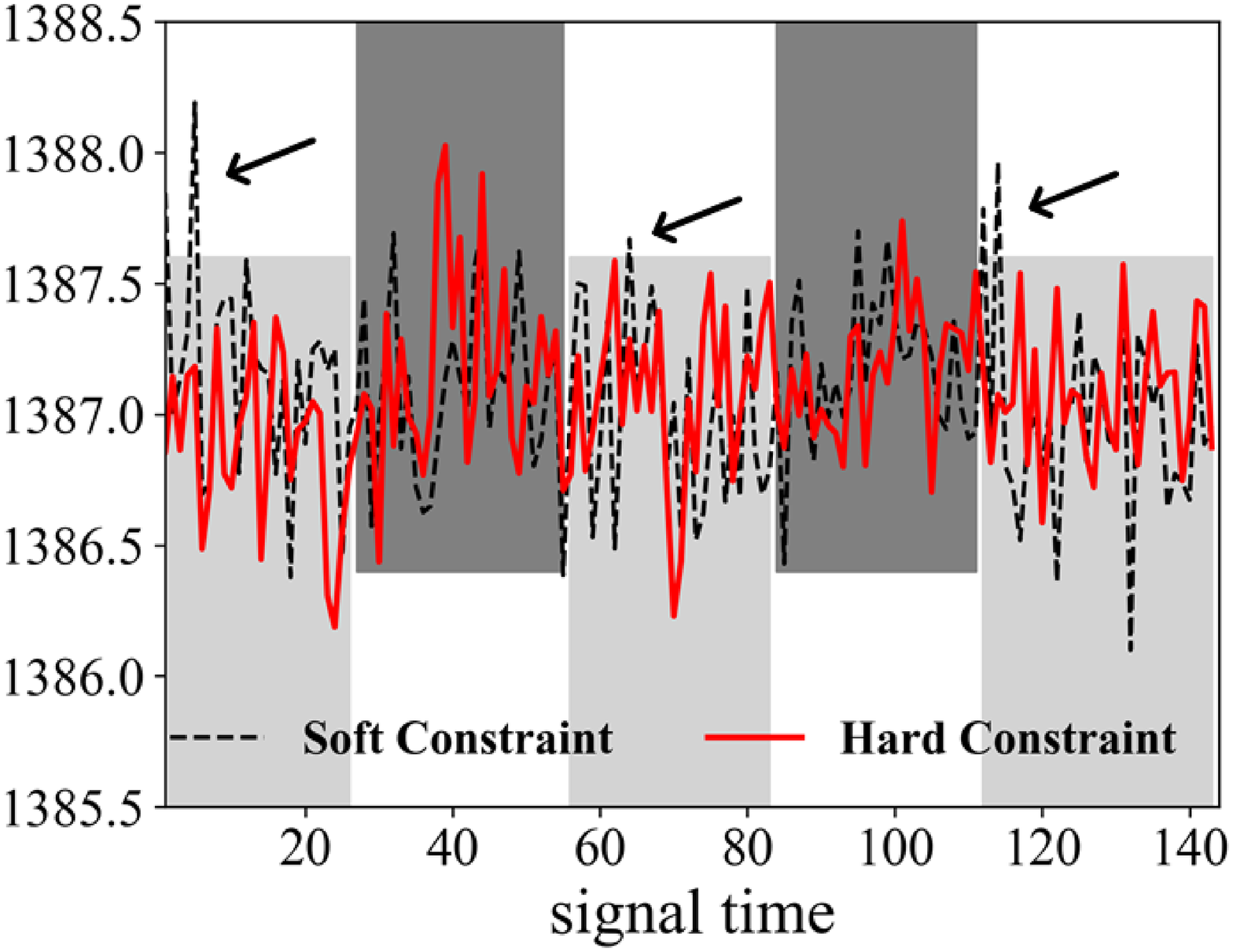}
                \label{fig:energy_level3}
        }
    \caption{Optimization under soft and three levels of hard constraints in the energy demand-response scenario. Light-grey shaded areas represent the upper bound and dark-grey shaded areas the lower bound. Arrows point to violations of hard constraints.}
    \label{fig:hard_energy}
\end{figure*}
\begin{figure*}
    \centering
    \subfigure[$\mathcal{U}_{u}=3$, $\mathcal{L}_{u}=-3$, $r=0.56$, $I=0.7220$.]{
                \includegraphics[height=2.8cm,width=4cm]{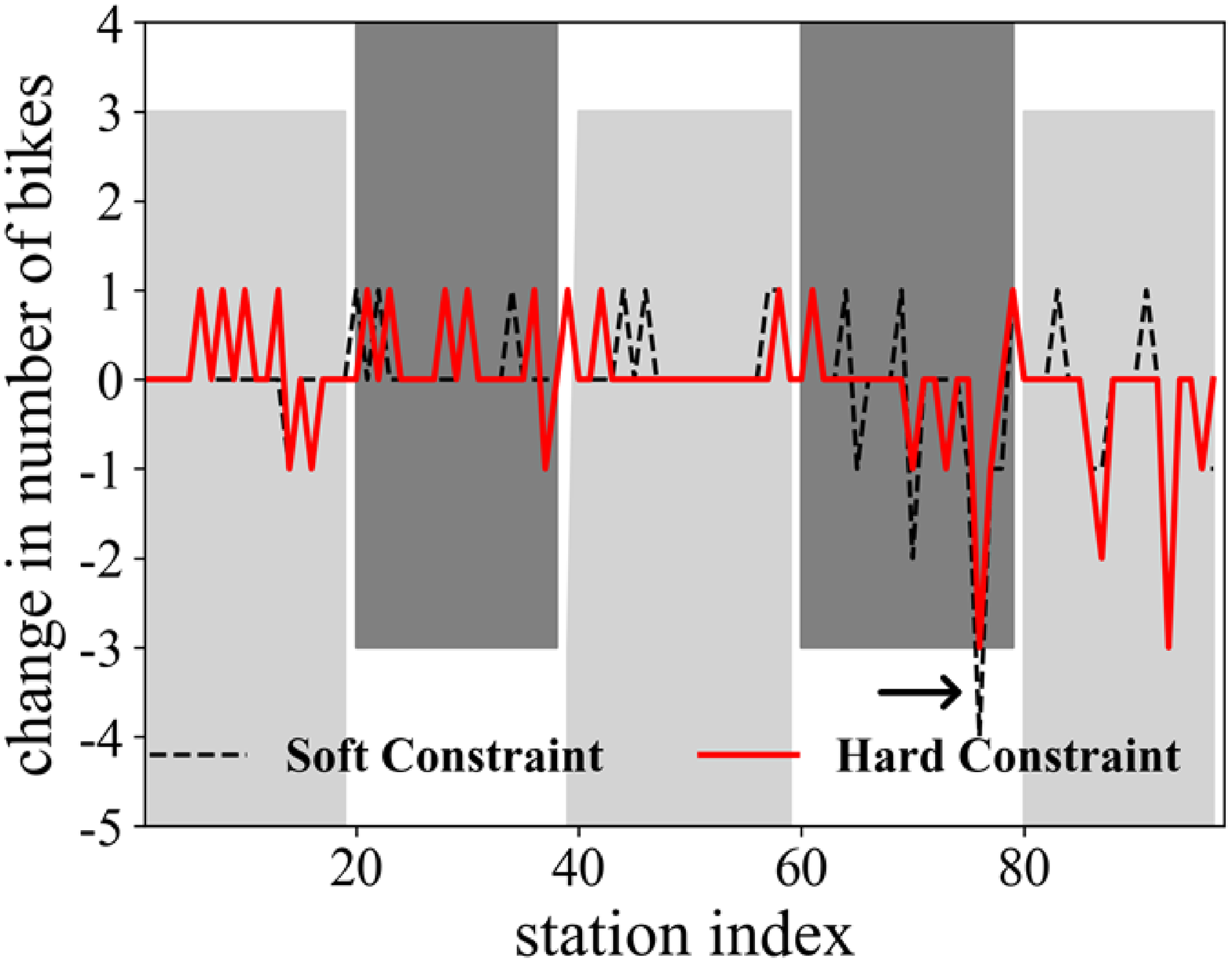}
                \label{fig:bicycle_level1}
        }
 \hspace{0.2cm}
    \subfigure[$\mathcal{U}_{u}=2$, $\mathcal{L}_{u}=-2$, $r=0.155$, $I=0.6793$.]{
                \includegraphics[height=2.8cm,width=4cm]{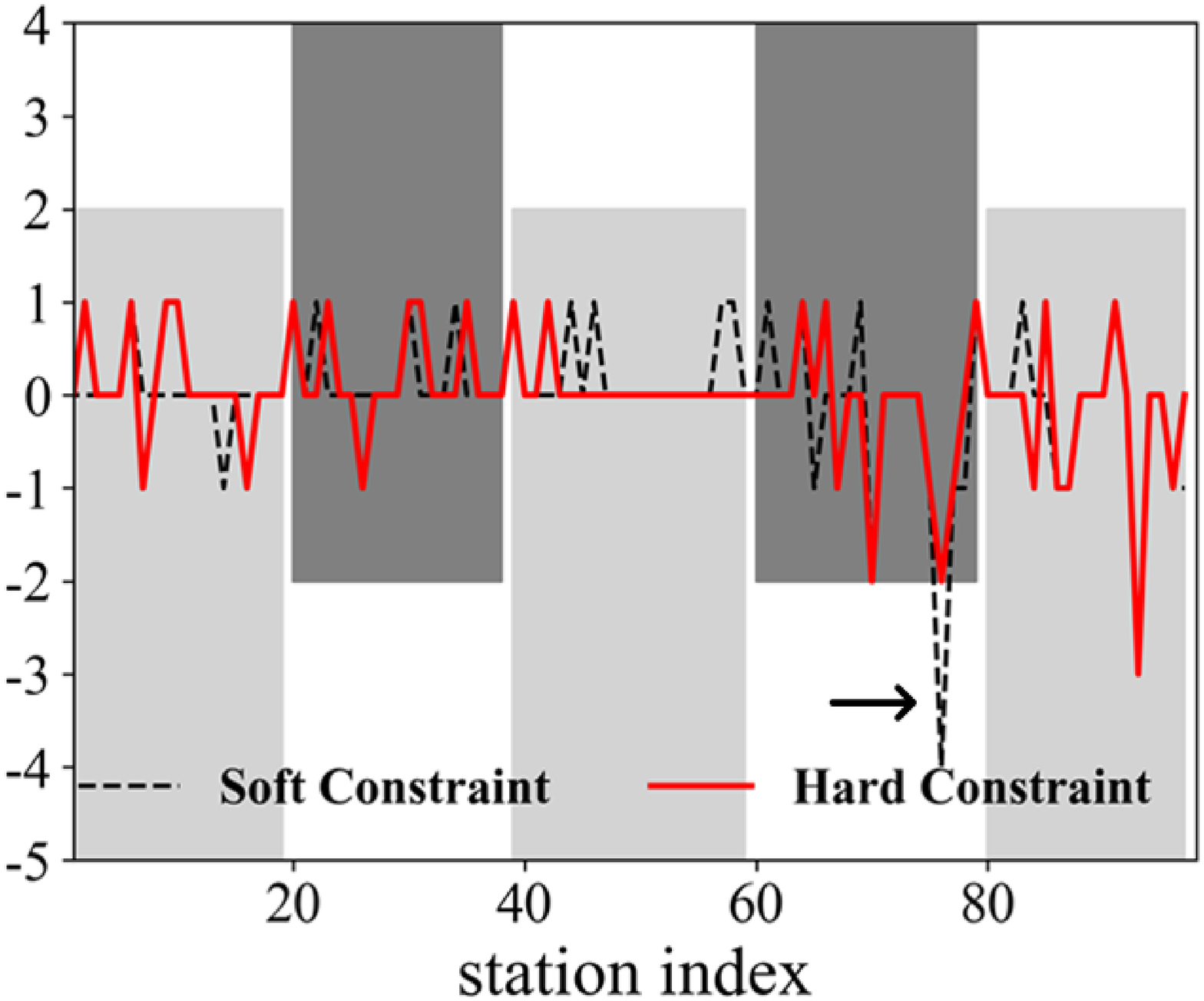}
                \label{fig:bicycle_level2}
        }
 \hspace{0.2cm}
    \subfigure[$\mathcal{U}_{u}=1$, $\mathcal{L}_{u}=-2$, $r=0.045$, $I=0.4328$.]{
                \includegraphics[height=2.8cm,width=4cm]{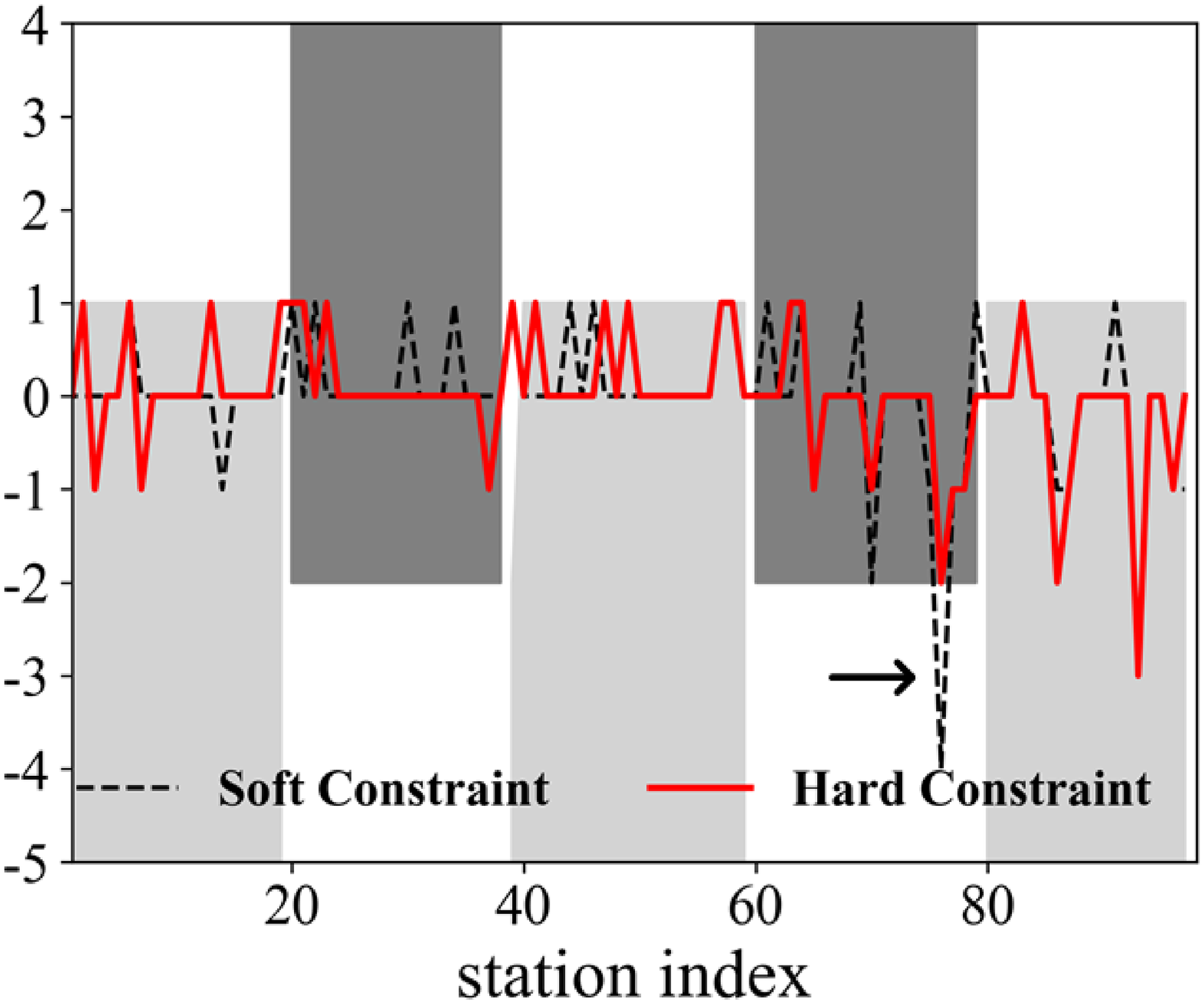}
                \label{fig:bicycle_level3}
        }
    \caption{Optimization under soft and three levels of hard constraints in the bike sharing scenario. Light-grey shaded areas represent the upper bound and dark-grey shaded areas the lower bound. Arrows point to violations of hard constraints.}
    \label{fig:hard_bicycle}
\end{figure*}

\begin{figure*}
    \centering
    \subfigure[$\mathcal{U}_{u}=800$, $\mathcal{L}_{u}=2200$, $r=1$, $I=7.9950$.]{
                \includegraphics[height=2.8cm,width=4cm]{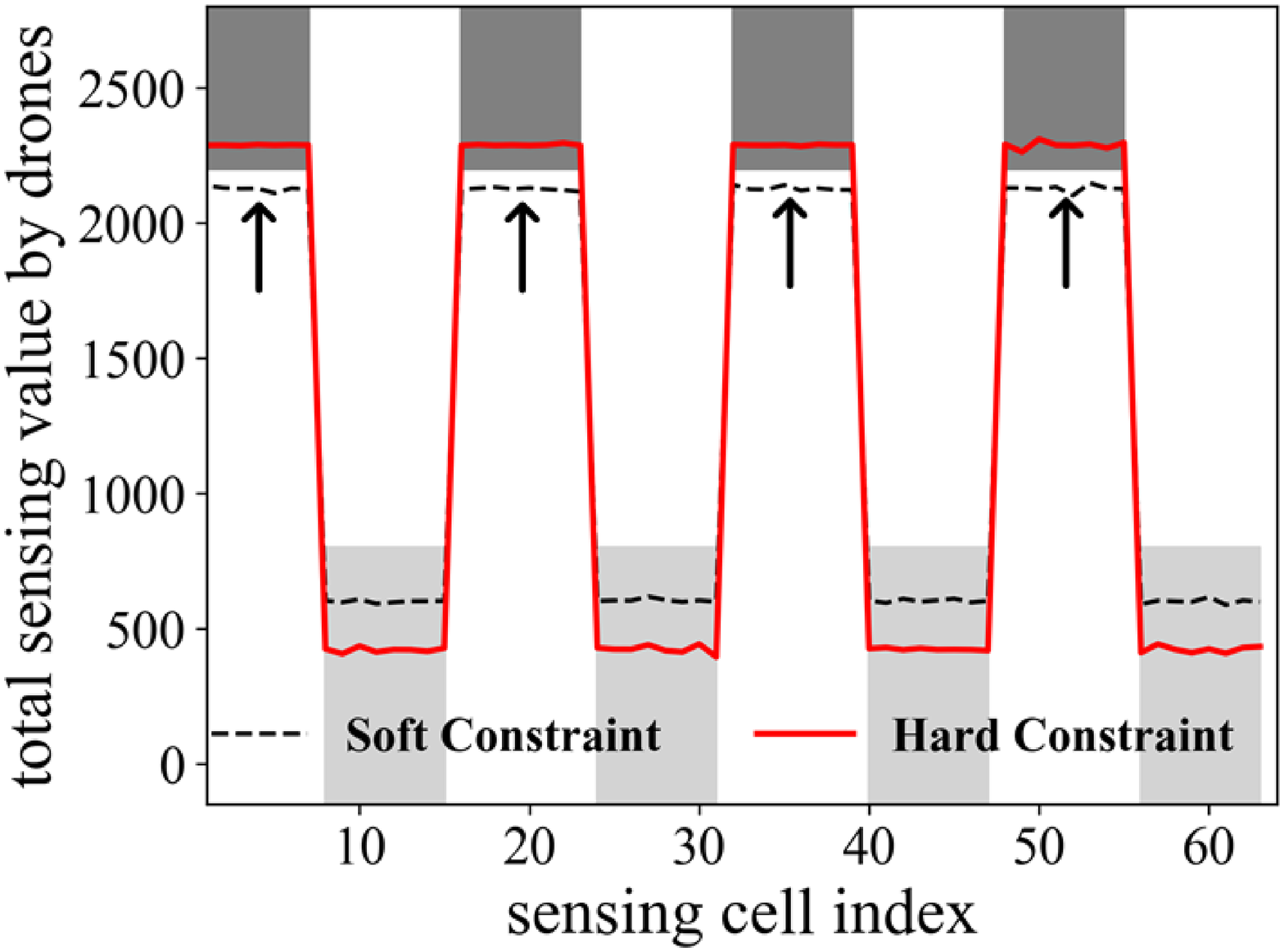}
                \label{fig:drone_level1}
        }
 \hspace{0.2cm}
    \subfigure[$\mathcal{U}_{u}=600$, $\mathcal{L}_{u}=2400$, $r=0.58$, $I=8.0561$.]{
                \includegraphics[height=2.8cm,width=4cm]{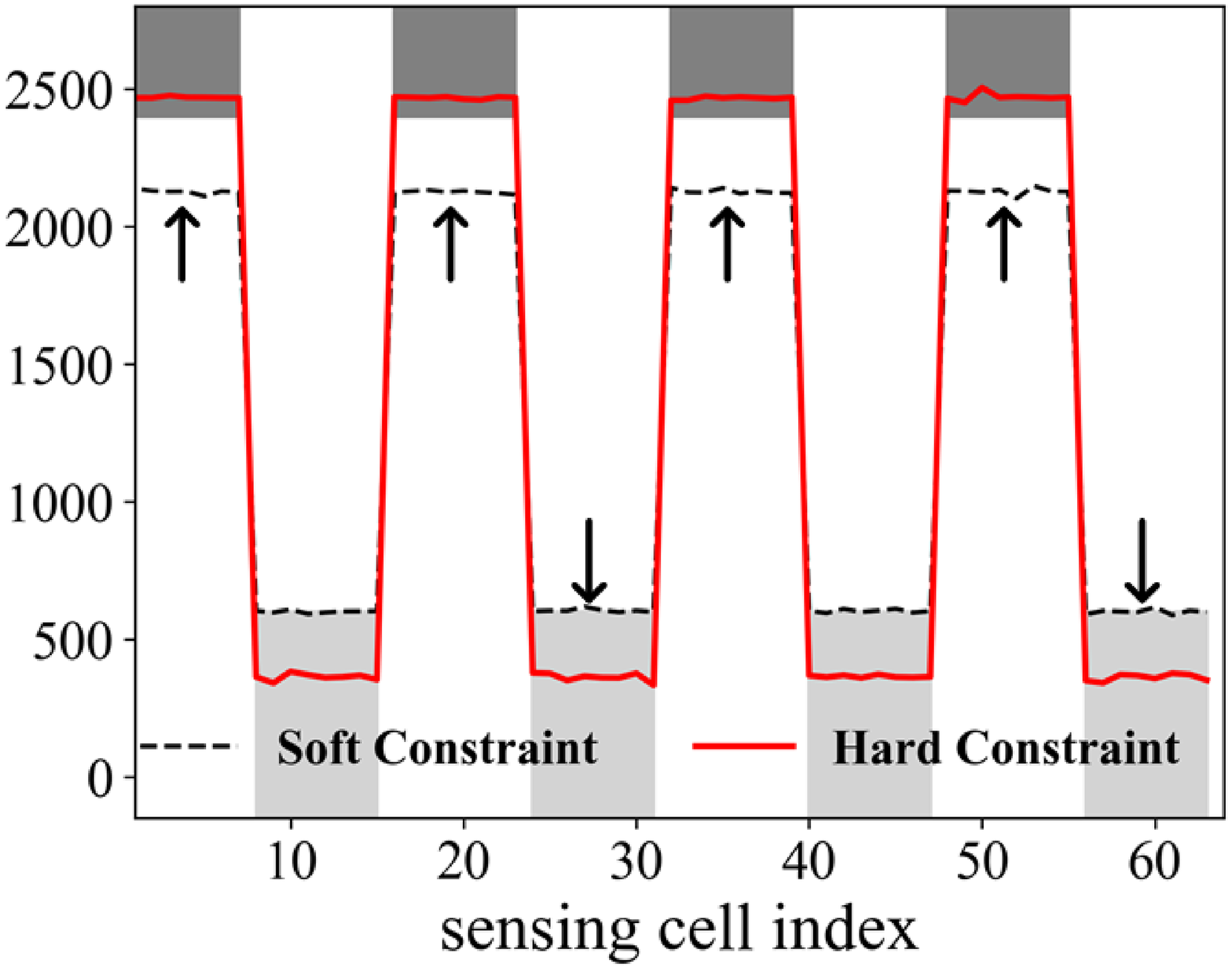}
                \label{fig:drone_level2}
        }
 \hspace{0.2cm}
    \subfigure[$\mathcal{U}_{u}=400$, $\mathcal{L}_{u}=2500$, $r=0.055$, $I=8.7810$.]{
                \includegraphics[height=2.8cm,width=4cm]{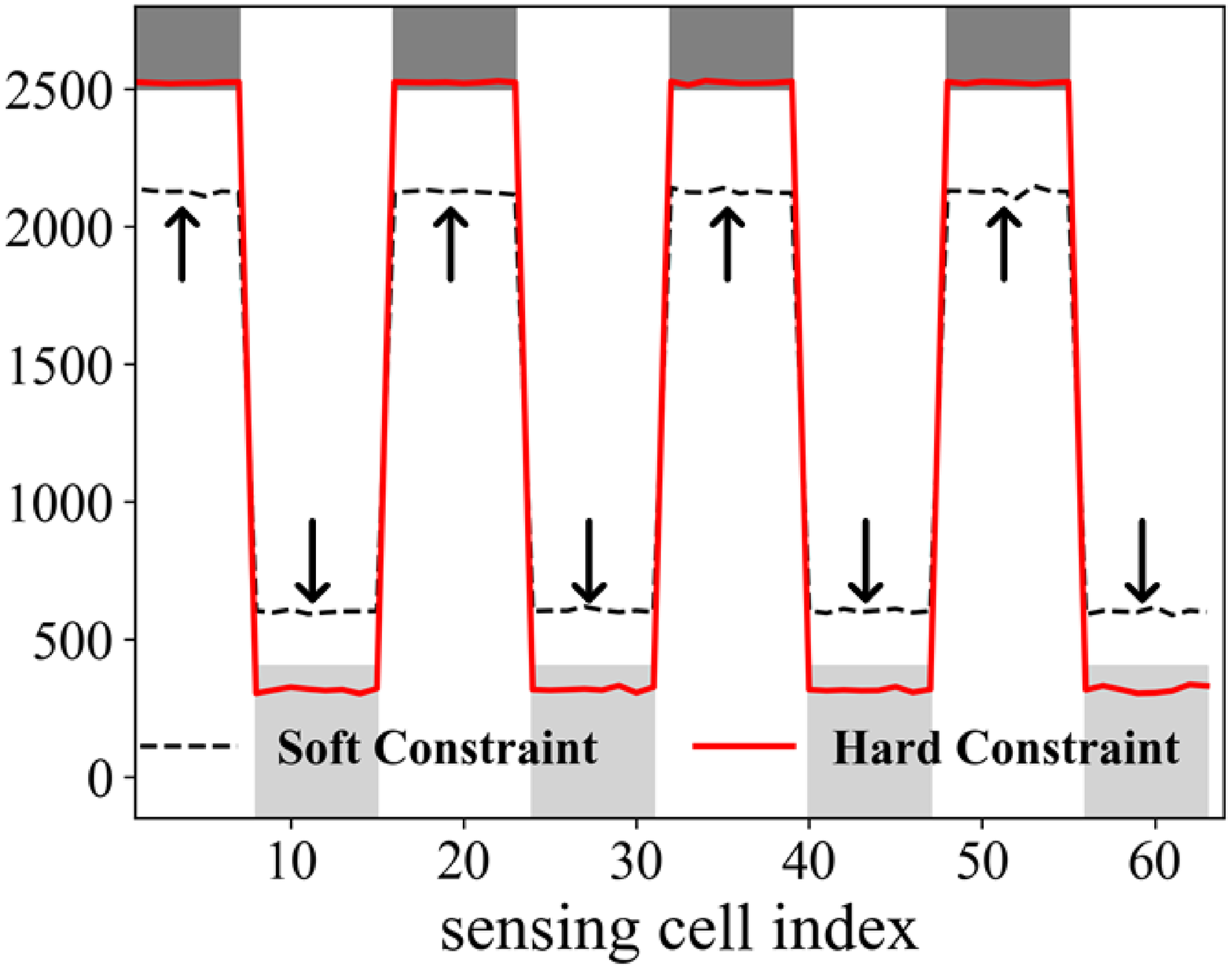}
                \label{fig:energy_level3}
        }
    \caption{Optimization under soft and three levels of hard constraints in the UAV swarm sensing scenario. Light-grey shaded areas represent the upper bound and dark-grey shaded areas the lower bound. Arrows point to violations of hard constraints.}
    \label{fig:hard_drone}
\end{figure*}

\cparagraph{Energy demand-response}
This dataset is based on energy disaggregation of the simulated zonal power transmission in the Pacific Northwest Smart Grid Demonstrations Project~\cite{pournaras2017self,Pournaras2023}. It contains 5600 consumers with their energy demand recorded every 5 min in a 12h span of a day. The goal is to perform power peak shaving to prevent blackouts~\cite{pournaras2018decentralized} by minimizing the variance.

\cparagraph{Bike sharing }
The Hubway Data Visualization Challenge dataset consists of the trip records of the Hubway bike sharing system in Paris~\cite{pournaras2018decentralized,Pournaras2023}. The data contain 2300 users, each with a varying number of possible plans for the bike stations from which bikes are picked up or returned (98 stations in total). The goal is to keep the bike sharing stations load balanced by minimizing the variance of the global plan. 

\cparagraph{UAV swarm sensing}
The dataset contains 1000 drones that can capture images or videos of vehicle traffic information on public roadways over 64 areas of interest (sensing cells) that are uniformly distributed in the city map~\cite{qin20223,Pournaras2023}. Drones aim to collect the required amount of sensing data (target plan) determined by a continuous kernel density estimation, for instance, monitoring cycling risk based on past bike accident data~\cite{castells2020cycling}.

\subsection{Hard constraint satisfaction works}

For each application scenario, three incremental levels of hard constraints are set to the aggregate choices (global plan $g$). These levels are quantiles chosen empirically by observing the median global plan after several executions of I-EPOS based on soft constraints. The agents are assumed here altruistic, such that: $\beta_{i}=0, \alpha_{i}=0, \forall i \in \{1,...,n\}$.

\begin{figure*}
    \centering
    \subfigure[Energy demand-response.]{
                \includegraphics[height=2.8cm,width=4.2cm]{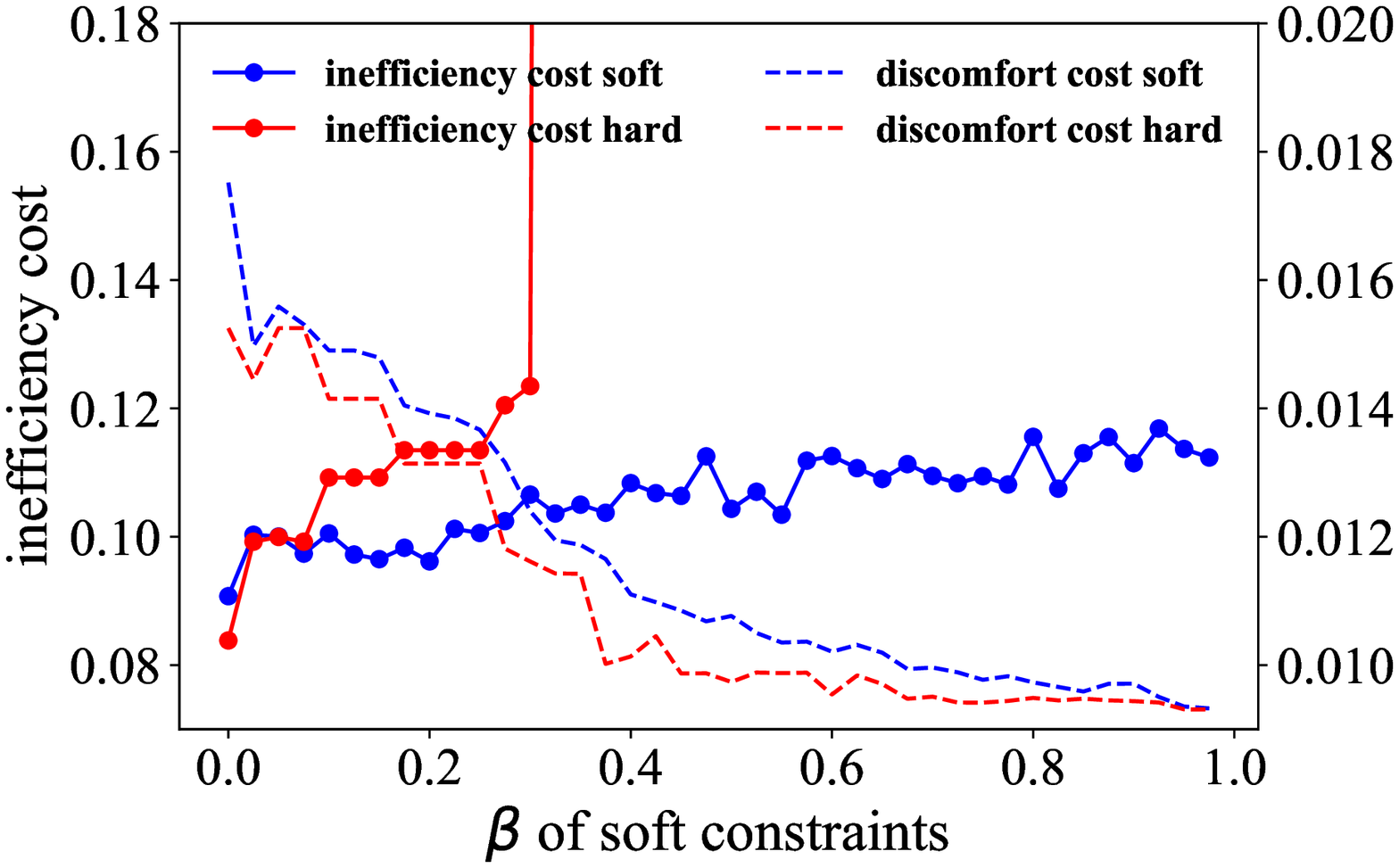}
                \label{fig:energy_local1}
        }
    \hspace{0.1cm}
    \subfigure[Bike sharing.]{
                \includegraphics[height=2.8cm,width=4.2cm]{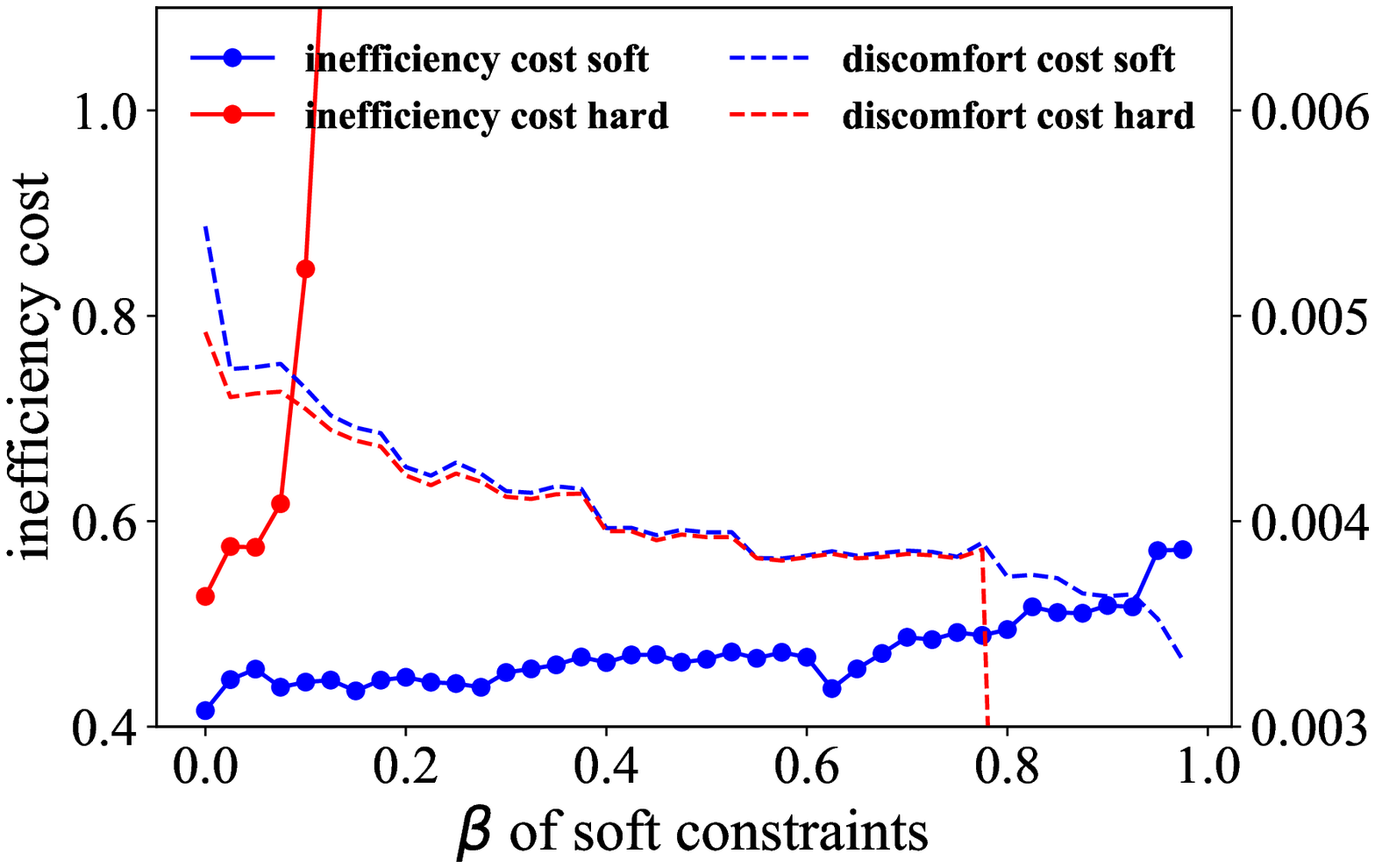}
                \label{fig:bicycle_local1}
        }
    \hspace{0.1cm}
    \subfigure[UAV swarm sensing.]{
                \includegraphics[height=2.8cm,width=4.2cm]{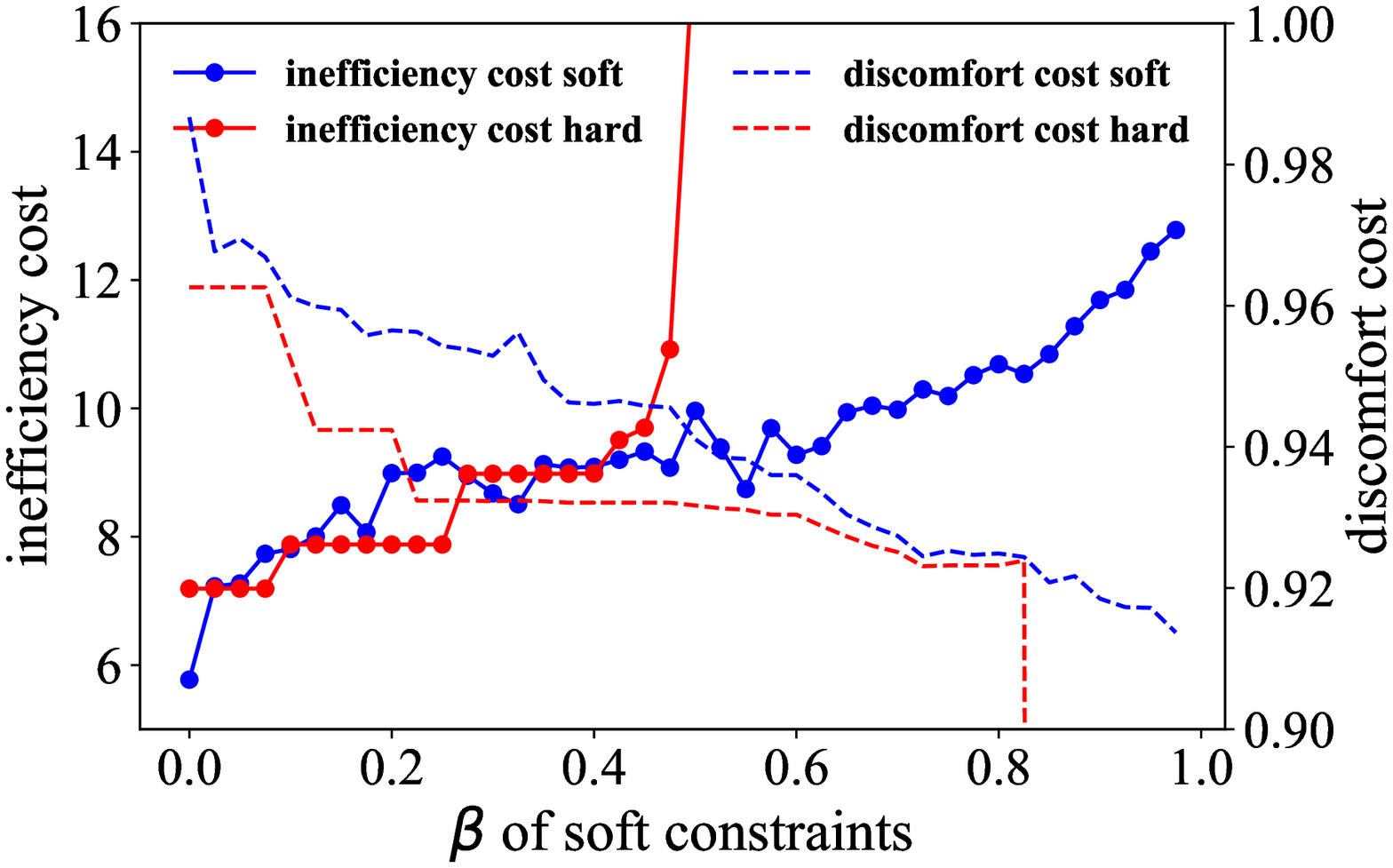}
                \label{fig:drone_local1}
        }
    \caption{Inefficiency and discomfort cost as a function of the altruism level for soft and hard constraints for the three Smart City application scenarios.}
    \label{fig:local1}
\end{figure*}

\begin{figure*}
    \centering
    \subfigure[Energy demand-response.]{
                \includegraphics[height=2.8cm,width=4.2cm]{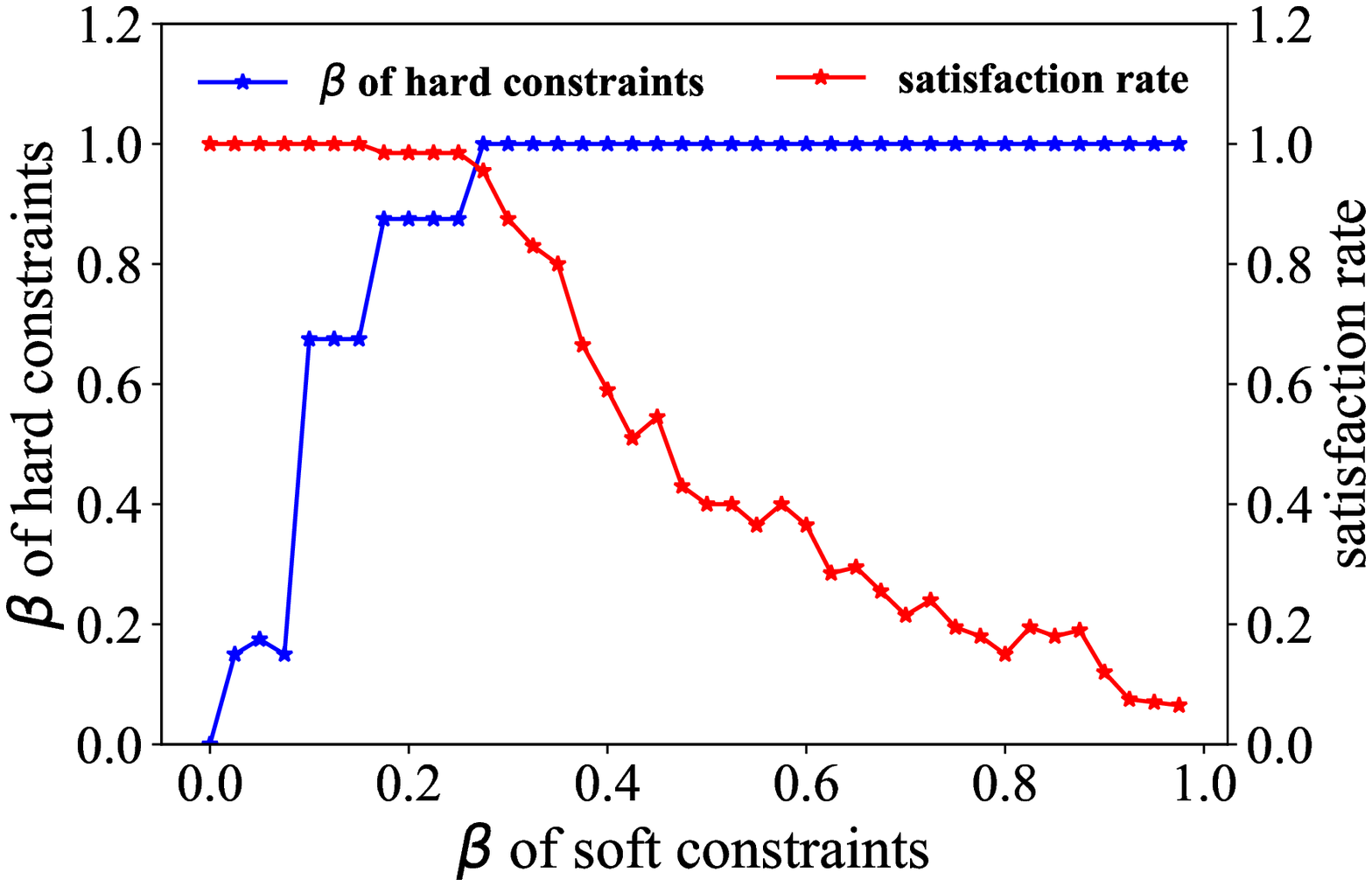}
                \label{fig:energy_local2}
        }
    \hspace{0.2cm}
    \subfigure[Bike sharing.]{
                \includegraphics[height=2.8cm,width=4.2cm]{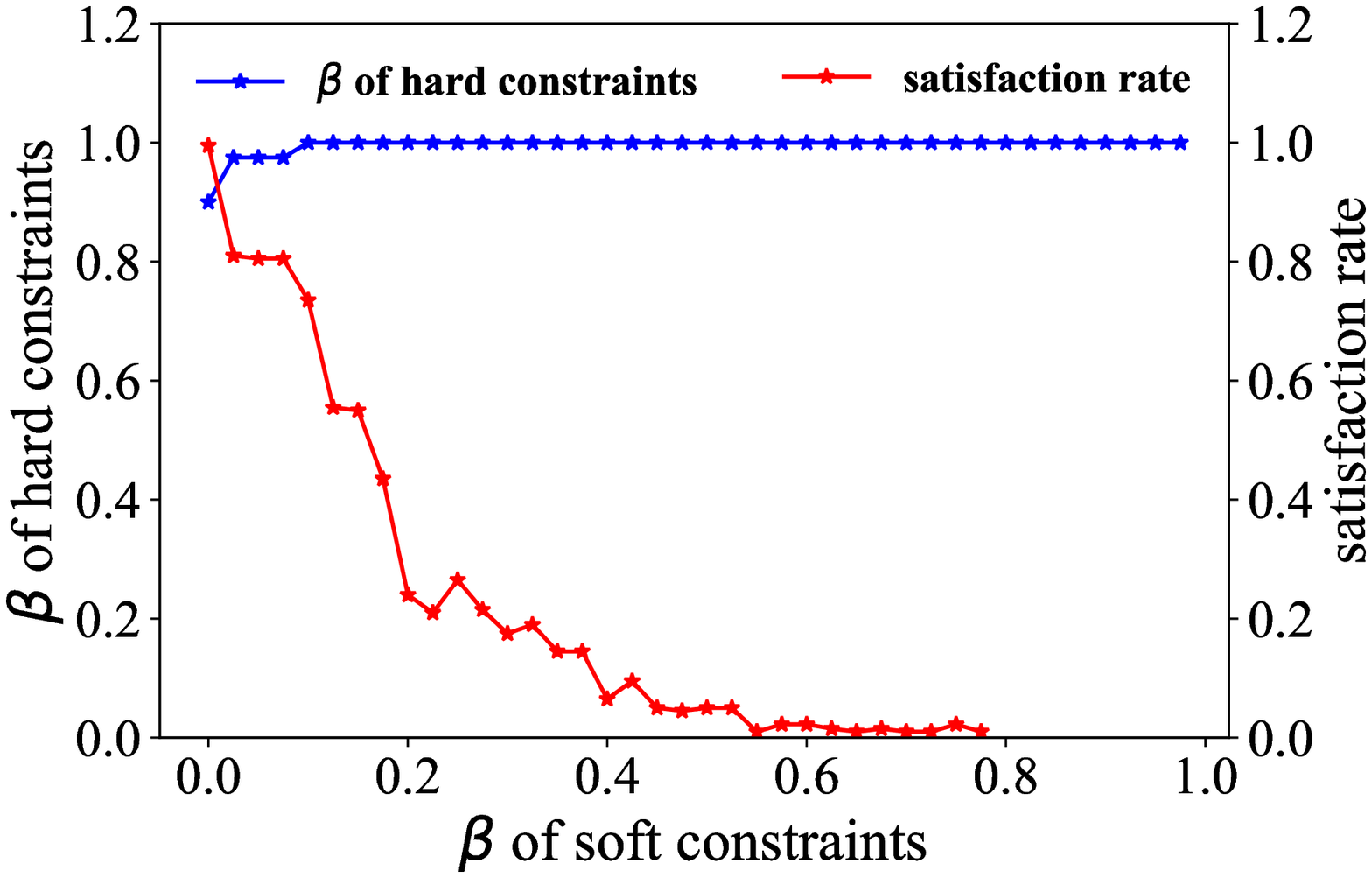}
                \label{fig:bicycle_local2}
        }
    \hspace{0.2cm}
    \subfigure[UAV swarm sensing.]{
                \includegraphics[height=2.8cm,width=4.2cm]{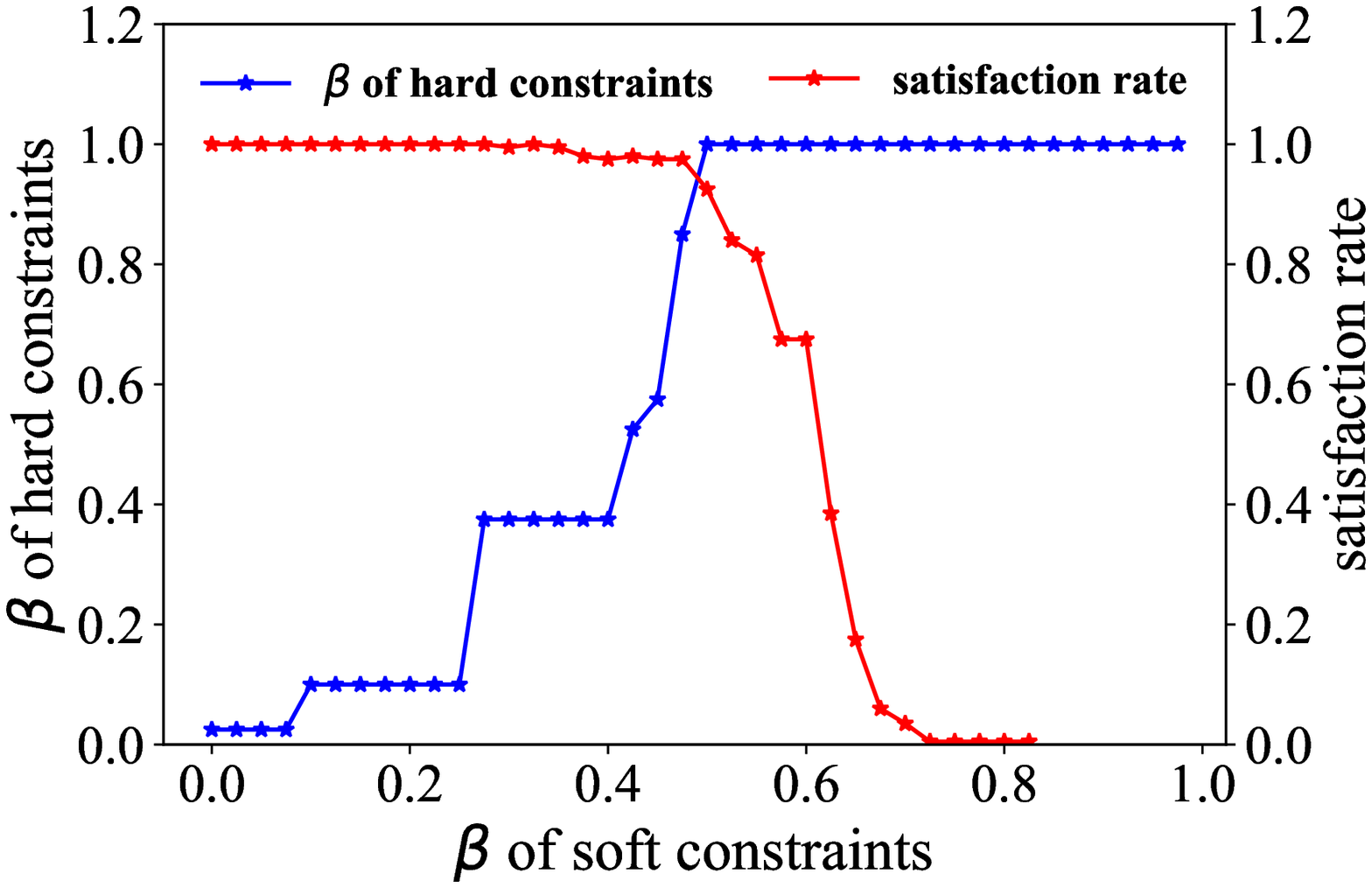}
                \label{fig:drone_local2}
        }
    \caption{Required behavioral shift to mitigate the performance degrade of satisfying hard constraints for the three Smart City application scenarios. The satisfaction rate is also shown for each scenario.
    Performance comparison: $\beta$ of soft constraints vs. $\beta$ of hard constraints and satisfaction rate.}
    \label{fig:local2}
\end{figure*}

In the scenario of energy demand-response, hard constraints based on upper and lower bounds may represent an envelope of operation within which demand does not cause a blackout. In the scenario of bike sharing, hard constraints may represent limits on incoming or outgoing bikes that infrastructure operators may have, e.g. parking capacity. In the scenario of UAV swarm sensing, an upper bound of hard constraints may represent privacy-sensitive areas or regulated no-fly zones for drones. In contrast, a lower bound may represent minimal information required to monitor effectively a phenomenon, e.g. a forest fire or traffic jam.

Figures~\ref{fig:hard_energy},~\ref{fig:hard_bicycle} and~\ref{fig:hard_drone} show the global plans for the soft constraint (baseline) along with three incremental and alternating levels of hard constraints (upper/lower bounds). Under soft constraints, the upper and lower bounds are violated, whereas hard constraints prevent these violations. Stricter hard constraints prevent more violations, however, the satisfaction rate ($r$) drops significantly, while the inefficiency cost increases. This shows that strict hard constraints are likely to oppose the soft constraints. Last but not least, note that the scenario of UAV swarm sensing allows the satisfaction of a larger number of hard constraints, while preserving high satisfaction rates. This is because the agents have more operational flexibility by generating a larger number of plans (64>24>10). 



\subsection{Behavior shift can mitigate hard constraints}

Satisfying hard constraints results in a degrade of the performance profile (lower inefficiency cost) achieved under soft constraints. The recovery from this degrade is measured here as the required social capital (behavioral shift) that agents need to offer such that soft and hard constraints have equivalent performance. The raise of social capital is measured by the reduction of the mean $\beta_{i}$ value in the population of agents that makes them more altruistic, see Equation~1. The following method is introduced to measure the behavioral shift: I)  Perform parameter sweep on I-EPOS under soft constraints for  $\beta_{i}=0$, to $\beta_{i}=1, \forall i \in \{1,...n\}$ with a step of 0.025. II)  For each I-EPOS execution in Step 1 with a $\beta_{i}$ value, a discomfort cost $D$ and an inefficiency cost $I$, run I-EPOS under a hard constraint on the mean discomfort cost with an upper bound value equals to $D$ (the one of the I-EPOS execution under soft constraints). III) Derive the increased inefficiency cost under the hard constraint on the discomfort cost. IV)  Find the $\beta_{i}$ value from Step 1 that has the closest inefficiency cost with the one derived in Step 3 under the hard constraint and V)  Compare the two $\beta_{i}$ values in Step 2 and 4. The difference represents the required mean behavioral shift to mitigate the performance degrade of hard constraints.

Figure~\ref{fig:local1} illustrates the inefficiency and discomfort cost as a function of $\beta_{i}$ under soft and hard constraints and the three different application scenarios. It becomes apparent that hard constraints require a minimum and significant level of altruism, otherwise, inefficiency cost rapidly explodes, especially in the scenarios without significant operational flexibility. This is also the reason why the discomfort cost becomes easier to reduce in the scenario of UAV swarm sensing, which comes with higher operational flexibility.

Figure~\ref{fig:local2} shows the required behavior shift to restore the performance loss as a result of satisfying hard constraints. For energy demand-response, the agents need on average $44.29\%$ higher altruism under hard constraints to meet the performance of the soft constraints. The bike sharing scenario suggests an almost complete shift from selfish to altruistic behavior. Strikingly, the scenario of UAV swarm sensing shows performance gain as a result of satisfying hard constraints. As the number of plans is significantly higher for the UAV dataset, the search space is larger, which affects the optimality of the collective iterative learning paradigm in I-EPOS. 

The satisfaction rate for the energy demand-response (Figure~\ref{fig:energy_local2}), bike sharing (Figure~\ref{fig:bicycle_local2}) and  UAV swarm sensing (Figure~\ref{fig:drone_local2})  are

\noindent
 54.45\%, 19.44\% and 61.21\% respectively. For $\beta \leq 0.475$ and $\beta \leq 0.25$, the satisfaction rate is 100\% for the UAV swarm sensing and energy demand-response respectively. The operational flexibility by higher number of possible plans increases the constraints satisfaction rate.


\section{Conclusion and Future Work}\label{sec:Conclusion}

To conclude, this paper shows that the decentralized satisfaction of global hard constraints is feasible. It is a significant enabler for sustainability and resilience in several Smart City application scenarios such as energy demand-response to avoid blackouts, load balancing of bike sharing stations to make more accessible low-carbon transport modalities as well as improved sensing quality and efficiency by swarms of energy-constrained drones. Results show that hard constraints can be easily violated when optimizing exclusively for soft constraints. Instead, the proposed model prevents to a very high extent such violations.

Results also reveal the performance cost when hard constraints are introduced and how this cost can be mitigated by a behavioral shift towards a more altruistic behavior that sacrifices individual comfort for collective efficiency. These findings are invaluable for informing policy makers and systems operators of the required social capital that they need to raise to satisfy ambitious policies such as net-zero.

The open-source software artifact implementation of the proposed model to the I-EPOS collective learning algorithm is a milestone to encourage further research and application scenarios based on decentralized hard constraint satisfaction. Future work includes the applicability of the model to other decentralized optimization algorithms. The proposed heuristic is designed to satisfy all hard constraints together, which may be a limitation for high numbers of such opposing constraints, i.e. sacrifice of optimality and low satisfaction rates. Instead, a more incremental (and possibly partial) satisfaction of the hard constraints is part of future work. The further understanding of how to recover missing social capital to preserve both efficiency and fairness is also subject of future 

\section*{Acknowledgements}\label{sec:Acknowledgment}

This work is supported by a UKRI Future Leaders Fellowship (MR\-/W009560\-/1): `\emph{Digitally Assisted Collective Governance of Smart City Commons--ARTIO}', the Alan Turing Fellowship project `\emph{New Edge-Cloud Infrastructure for Distributed Intelligent Computing}' and the SNF NRP77 `Digital Transformation' project "Digital Democracy: Innovations in Decision-making Processes", \#407740\_187249.

\bibliographystyle{ACM-Reference-Format}
\bibliography{sample}


\end{document}